\documentclass[12pt,a4paper,final]{iopart}

\usepackage{iopams}  
\usepackage{graphicx}
\usepackage[breaklinks=true,colorlinks=true,linkcolor=blue,urlcolor=blue,citecolor=blue]{hyperref}
\begin{document}

\title[]{Edge channel confinement in a bilayer graphene $n$-$p$-$n$ quantum dot}

\author{Hiske Overweg, Peter Rickhaus, Marius Eich, Yongjin Lee, Riccardo Pisoni, Thomas Ihn, Klaus Ensslin}
\address{Solid State Physics Laboratory, ETH Z{\"u}rich, CH-8093 Z{\"u}rich, Switzerland}

\ead{overwegh@phys.ethz.ch}

\author{Kenji Watanabe, Takashi Taniguchi}
\address{National Institute for Material Science, 1-1 Namiki, Tsukuba 305-0044, Japan}

\begin{abstract}
We combine electrostatic and magnetic confinement to define a quantum dot in bilayer graphene. The employed geometry couples $n$-doped reservoirs to a $p$-doped dot. At magnetic field values around $B = 2~$T, Coulomb blockade is observed. This demonstrates that the coupling of the copropagating modes at the $p$-$n$ interface is weak enough to form a tunnel barrier, facilitating transport of single charge carriers onto the dot. This result may be of use for quantum Hall interferometry experiments.
\end{abstract}


\section{Introduction}

Graphene is thought to be an attractive host material for spin qubits, because charge carriers in graphene are predicted to have long spin coherence times due to the small spin-orbit and hyperfine interaction  \cite{Trauzettel2007,Fuchs2012,Droth2013,Fuchs2013}. So far quantum dots in graphene, studied from a theoretical perspective in Refs.~\cite{Pereira2007,Recher2009,Guclu2014}, have been realized by etching \cite{Ponomarenko2008,Stampfer2008}, electrostatic confinement by lithographic gates \cite{Allen2012,Goossens2012a} or STM tips \cite{Lee2016a,Freitag2016a}, and exploiting the disorder potential \cite{Martin2009,Tovari2016b} and have only been studied in the unipolar regime. The ambipolar nature of graphene is widely viewed as a hurdle to overcome when trying to confine charge carriers in graphene. Because of the absence of a band gap, it is not possible to deplete graphene locally by applying an appropriate gate voltage, as is the case for GaAs \cite{Zheng1986,Thornton1986} and other two-dimensional semiconducting materials \cite{Chou2005,Song2015}.

Bilayer graphene can be used to solve this problem. Pristine bilayer graphene has a gapless parabolic bandstructure, as shown by the dashed green line in Fig.~\ref{fig:0}a. However, a displacement field leads to the opening of a band gap \cite{McCann2006a,McCann2007,McCann2007a,Oostinga2008,Zou2010,Taychatanapat2010}, as indicated by the blue line in Fig.~\ref{fig:0}a. In a device with a  top- and back gate, the charge carrier density $n$ and displacement field $D$ can be tuned separately. The charge carrier density is given by $n =  (C_\mathrm{BG} V_\mathrm{BG}+ C_\mathrm{TG} V_\mathrm{TG})/e$ and the displacement field is given by $D = (C_\mathrm{BG} V_\mathrm{BG}- C_\mathrm{TG} V_\mathrm{TG})/\epsilon_0 $, where $e$ is the electron charge, $\epsilon_0$ the vacuum dielectric constant and $C_i$ the capacitance per unit area of gate $i$. We recently demonstrated that by using a graphite back gate and a metal top gate, leakage resistance under the  top gate of $10~$G$\Omega$ can be achieved in a displacement field of $D = 0.7~$V/nm \cite{Overweg2017}.

The tunability of the bandstructure can be exploited to define a nanostructure: by opening a band gap under the gates and tuning the gate voltages such that the Fermi level lies in the band gap, flow of charge carriers under the gates can be obstructed. In this work we use a split gate structure to define a narrow channel for current flow. An extra gate on top of the channel allows us to couple electron-like edge channels in the bulk of the device to a hole-like region in the channel, thus defining a $p$-doped quantum dot with $n$-doped reservoirs. We show the pinch-off characteristics of the quantum dot, report on the observation of Coulomb peaks and discuss similarities with quantum dots in GaAs, where the influence of a quantizing magnetic field has been extensively studied \cite{Wees1989, Taylor1992, Staring1992,Heinzel1994,VanderVaart1997,Zhang2009a,Baer2013}.

\section{Experimental details}
\begin{figure*}
\centering
\includegraphics[width=1\textwidth]{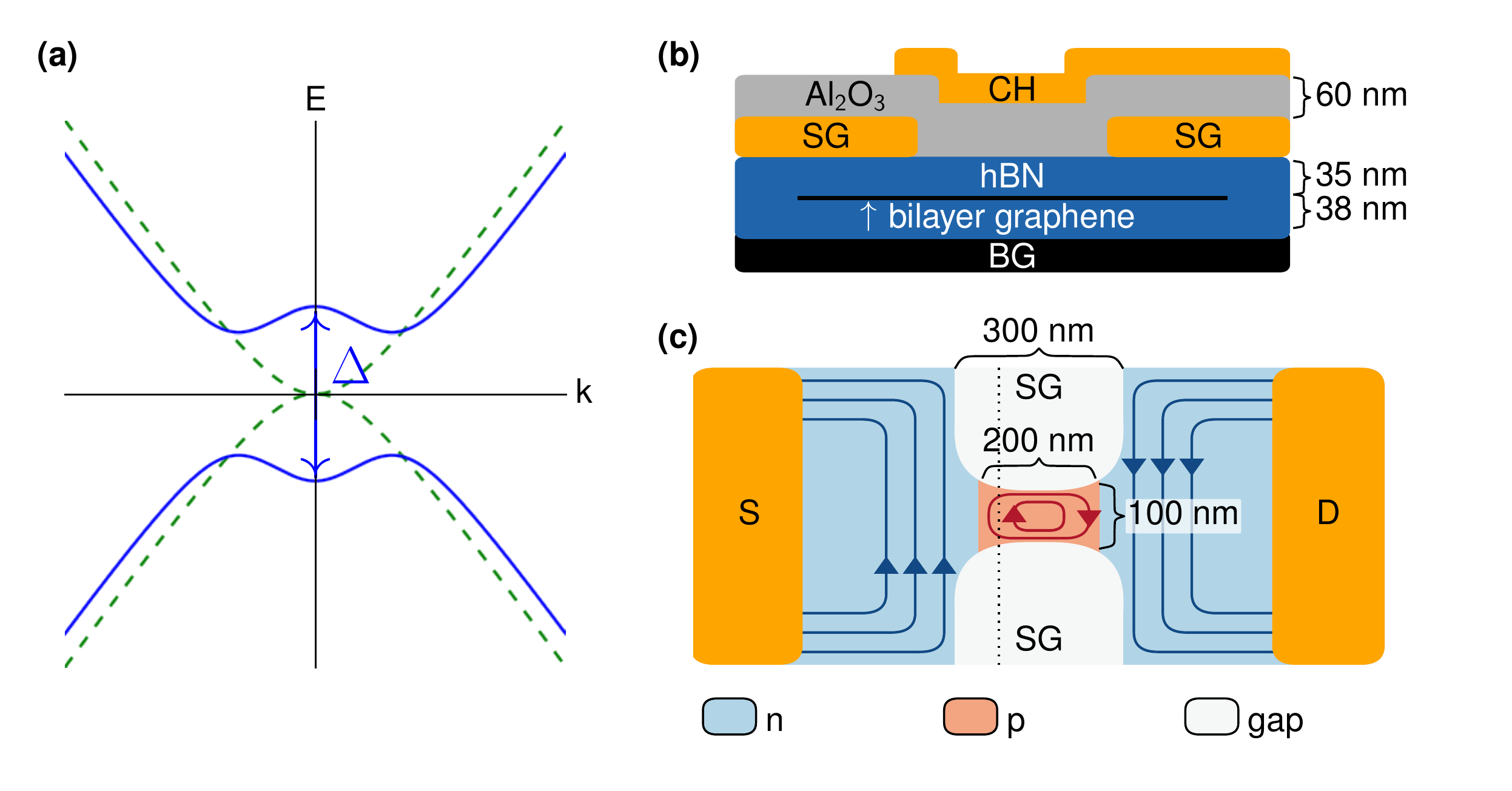}
\caption{(a) Bilayer graphene dispersion relation in presence (blue line) and absence (green line) of a displacement field (b) Side view of the sample. A bilayer graphene flake is encapsulated in hexagonal boron nitride (hBN). It has a graphite back gate below (BG), two split gates (SG) and a channel gate (CH) on top. The channel gate is separated from the split gates by a dielectric layer of Al$_2$O$_3$. (c) Top view of the sample. The graphene is contacted by a source (S) and a drain (D). It can be tuned to the regime where the bulk of the sample is in the $n$-regime, the channel is in the $p$-regime and the region under the split gates is depleted. When applying a perpendicular magnetic field, this leads to confined edge channels in the constriction. The dotted line indicates the position of the cross section shown in (b).}
\label{fig:0}
\end{figure*}

A cross section of the sample is shown in Fig.~\ref{fig:0}b. A bilayer graphene flake was encapsulated in hexagonal boron nitride and deposited on a graphite back gate (BG) on a Si/SiO$_2$ substrate. The graphene and graphite were contacted using the method described by Wang et al. \cite{Wang2013}. On top of the device, two split gates (SG) with a width of 300 nm were evaporated. After the deposition of a layer of Al$_2$O$_3$, the device was finalized by the deposition of a gate on top of the 100 nm wide channel between the split gates, the so-called channel gate (CH). The channel gate has a width of 200 nm. More details can be found in Ref.~\cite{Overweg2017}.
By adjusting all gate voltages carefully, the sample can be tuned to a regime where the region under the split gates is depleted, the bulk of the sample is $n$-doped (denoted by the blue color) and the channel is $p$-doped (red color). In the quantum Hall regime this leads to copropagating edge channels (see Fig.~\ref{fig:0}c), a topic widely studied at the moment \cite{Williams2007,Ozyilmaz2007,Amet2014,Morikawa2015,Zimmermann2016a,Wei2017a}. The edge channels inside the electrostatically defined channel in the device investigated in this work are confined to a small region in space. We observe that the electron-like and hole-like edge channels couple only weakly and thus the $p$-$n$ interfaces define tunnel barriers. As a result, the edge channels inside form a quantum dot.

All measurements were performed in a dilution refrigerator at a base temperature of 130~mK. A constant ac voltage bias was applied using conventional lockin techniques.

\section{Results and discussions}

\subsection{Characterization measurements}

\begin{figure*}
\centering
\includegraphics[width=0.9\textwidth]{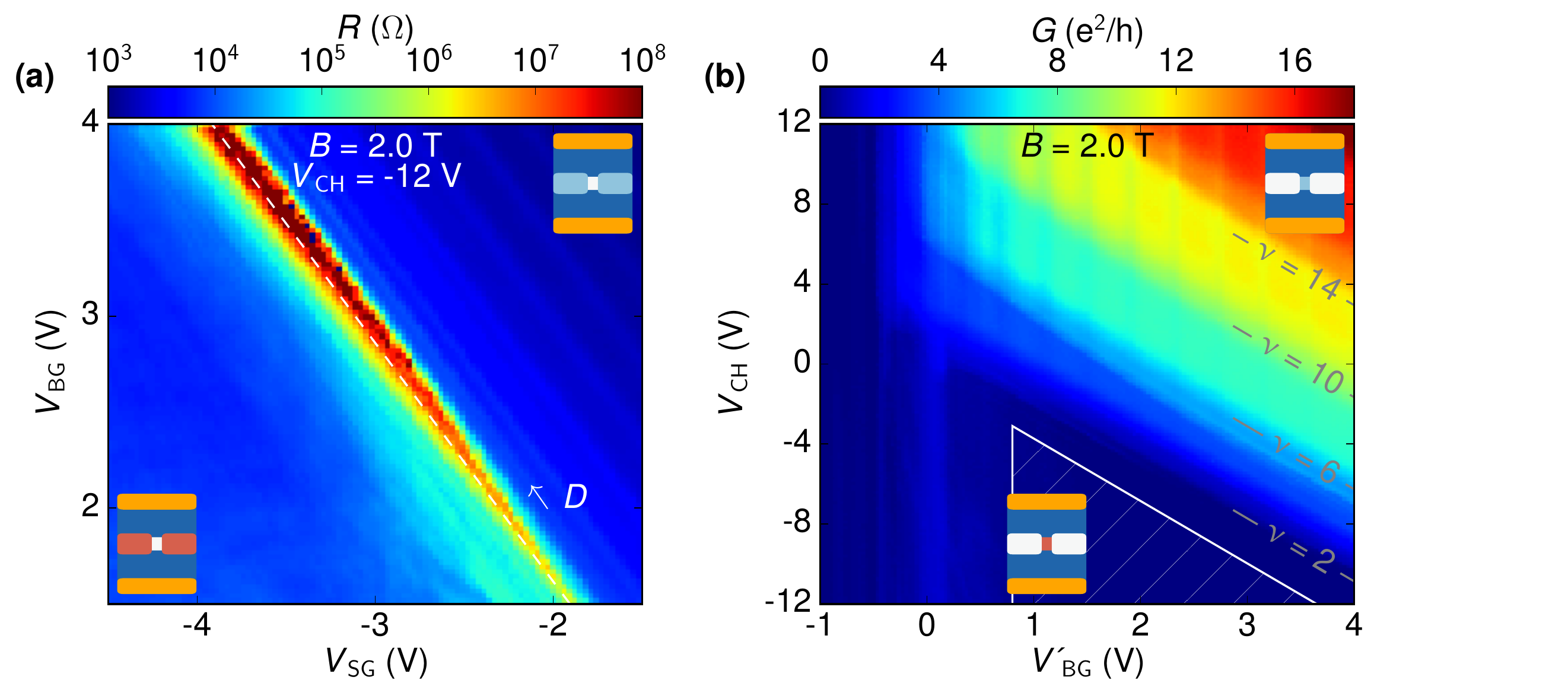}
\caption{ Sample characterization. (a) Resistance as a function of split gate and back gate voltage at $B~=~2~$T. Along the diagonal line of high resistance, the Fermi level underneath the split gates and in the channel is in the band gap. The resistance in the gap is higher than 1000 h/e$^2$. (b) Conductance at $B~=~2~$T as a function of channel gate voltage and back gate voltage. The split gate voltage is adjusted so as to keep the region under the split gates depleted (white dashed line in (a)). Conductance larger than e$^2$/h occurs when both the bulk and the channel are $n$-doped. Low conductance occurs when the bulk is $n$-doped and the channel is $p$-doped. Within the hatched area the edge channels inside the channel can form a quantum dot, as shown schematically in Fig.~\ref{fig:0}b.}
\label{fig:1}
\end{figure*}

To deplete the region under the split gates, we apply a positive voltage to the back gate and a negative voltage to the split gates (or vice versa). This leads to an asymmetry between the two graphene layers, which results in the opening of a band gap \cite{McCann2007a}. A detailed characterization of the band gap of this device at $B~=~0~$T can be found in Ref.~\cite{Overweg2017}. To find the strongest depletion of the region under the split gates in a finite magnetic field, we measure the resistance as a function of split gate voltage $V_\mathrm{SG}$ and back gate voltage $V_\mathrm{BG}$ at $B = 2~$T (Fig.~\ref{fig:1}a) and identify the line of highest resistance (white dashed line). For this measurement the channel gate voltage was fixed at a large negative value ($V_\mathrm{CH} = - 12~$V). With this channel gate voltage setting, the channel is depleted or $p$-doped for the entire range of the measurement and has a small coupling to the $n$-doped reservoirs (see also Fig.~\ref{fig:1}b). With a highly resistive channel and a highly $n$-doped bulk, the changes in resistance observed in this measurement stem primarily from the region under the split gates (see insets). If multiple regions of the device had a considerable influence on the resistance, features with different slopes would be expected, because of different gate capacitances in different regions. The diagonal line of high resistance corresponds to the charge neutrality point under the split gates. Along this line the displacement field increases in the direction of the arrow. We observe an increase of the resistance by two orders of magnitude along this line, ensuing from the increasing band gap under the split gates. The high resistance achieved is important for the formation of a well defined quantum dot. In all following measurements the split gate voltage is adjusted when the back gate voltage $V_\mathrm{BG}$ is varied (see white dashed line in Fig.~\ref{fig:1}a), so as to keep the region under the split gates depleted. We denote this by $V'_\mathrm{BG}$.

We now focus on the effect of the channel gate. The conductance at $B = 2~$T as a function of $V'_\mathrm{BG}$ and $V_\mathrm{CH}$ is shown in Fig.~\ref{fig:1}b. In the upper right corner, both the bulk and the channel are $n$-doped and therefore quantized conductance is observed, similar to Refs.~\cite{Williams2007,Ozyilmaz2007}. The observed lines with negative slope are defined by a constant filling factor in the channel. The transitions between the Landau levels are marked by their respective filling factors. They are used to determine the dependence of the charge carrier density on the gate voltages. The conductance inside the channel reaches $G = 16~$e$^2$/h, which shows that the edge channels of at least four Landau levels fit inside the channel. The extent of the wave function of the fourth Landau level is $l_B \sqrt{2N+1} = 55~$nm, where $l_B$ is the magnetic length. This extent is indeed smaller than the lithographic width of the channel.
In the hatched area at the bottom of the figure the conductance is much lower. In this regime, where the channel is $p$-doped while the bulk is $n$-doped, a quantum dot as sketched in Fig.~\ref{fig:0}c forms.

\subsection{Coulomb blockade measurements}

\begin{figure*}
\centering
\includegraphics[width=0.9\textwidth]{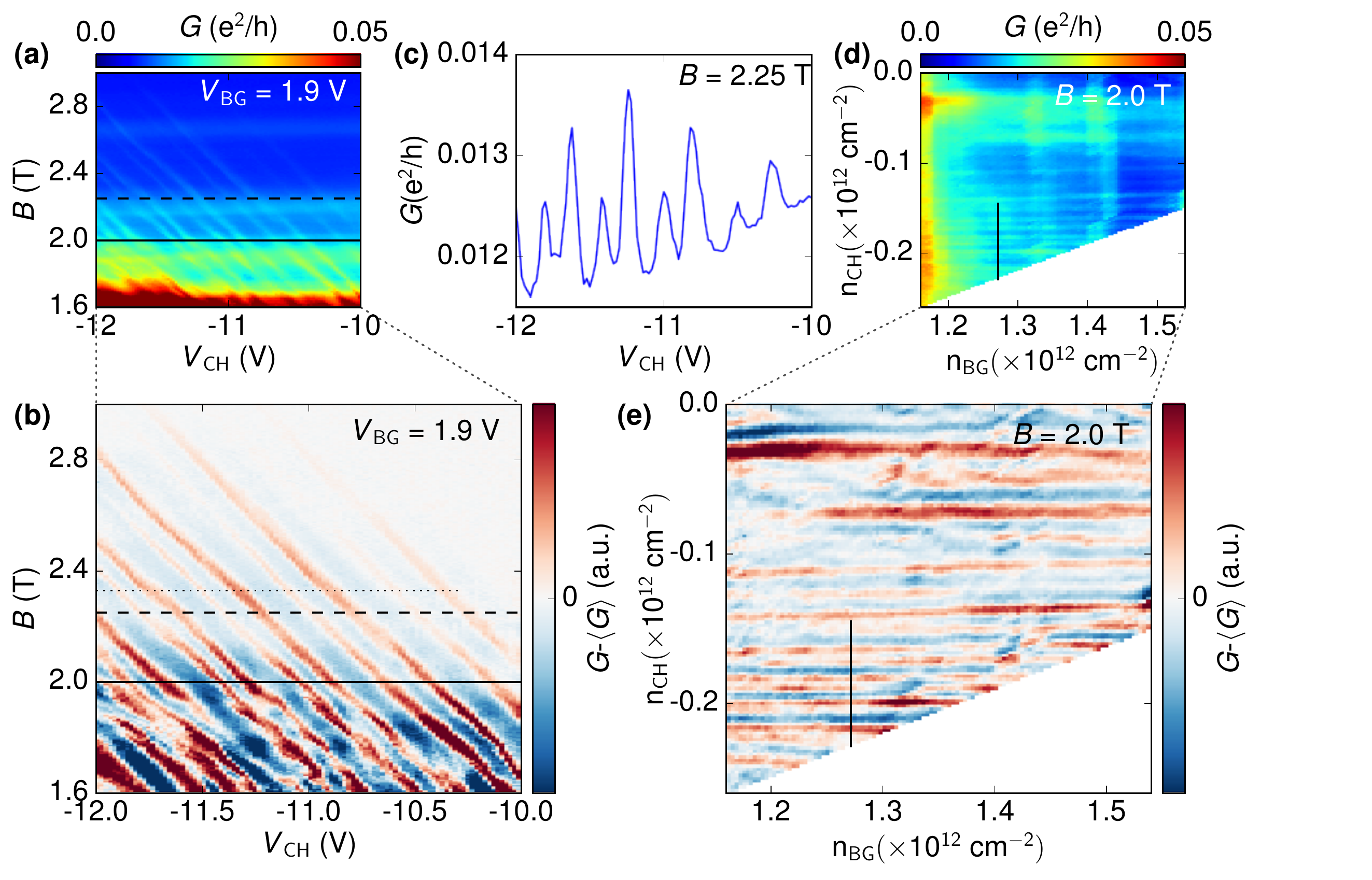}
\caption{Magnetotransport through the device. (a) Conductance as a function of channel gate voltage and magnetic field. Peaks corresponding to charging of the quantum dot show up as diagonal lines. To enhance the visibility, we subtract a smoothened background  (b). (c) Conductance at $B~=~2.25~$T as a function of channel gate voltage (dashed line in (a),(b)), showing an alternating sequence of low and high peaks. (d) Conductance at $B~=~2~$T as a function of the filling factors in the bulk and in the channel. The peaks in the conductance are marked by dashed lines. (e) Same as (d) with a smoothened background subtracted. The solid line in (b) corresponds to the solid line in (e).}
\label{fig:2}
\end{figure*}

Figure \ref{fig:2} shows the main result of this work. In Fig.~\ref{fig:2}a the conductance as a function of channel gate voltage and magnetic field is shown. For this measurement a dc bias of $V_\mathrm{SD,DC} = 200~\mu$V was applied to enhance the signal to noise ratio. Lines of higher conductance with a negative slope are observed. To enhance the visibility, we subtract a smoothened background (Fig.~\ref{fig:2}b). When moving in the direction of increasingly negative channel gate voltage, we can interpret each resonance as the addition of an individual charge carrier to the confined area. The negative slope implies that an increase of the magnetic field leads to the removal of charge carriers from the dot. This trend can be understood as follows: when the magnetic field increases, the hole-like Landau levels shift down in energy. Since the Fermi level in the dot is pinned by the reservoirs, this leads to a decrease of the number of occupied states in the dot. In the case of Aharonov-Bohm interferometry, which typically requires smaller dots \cite{Zhang2009a}, the opposite slope would be expected: an increase in magnetic field increases the flux through the interferometer, which has the same effect as increasing the area by applying a more negative gate voltage.

Charging lines of quantum dots in a magnetic field often show a slope related to a certain filling factor \cite{Ilani2004,Rosenow2007}. The slope of the lines in Fig.~\ref{fig:2}b is close to a filling factor of four inside the channel, but the error bar on the density inside the channel does not allow for a quantitative comparison. To calculate the density, we extracted the capacitance between the conducting channel and the channel gate $C_\mathrm{CH}$ from a Landau fan measured in a previous cool down (see Ref.~\cite{Overweg2017}) using a plate capacitor model. We also extract the same capacitance from several maps of the conductance at fixed magnetic field, such as Fig.~\ref{fig:1}b. The numbers we find are off by 30\% and we therefore conclude that our estimates of charge carrier density and filling factor inside the channel have an error bar of 30\%. Because of the vicinity of the split gates to the channel, it could be the case that $C_\mathrm{CH}$ depends on the split gate voltage. Moreover, $C_\mathrm{CH}$ might depend on the extent of the wave function inside the channel. These two factors are not accounted for in a simple capacitance model.

In Fig.~\ref{fig:2}c a line cut of Fig.~\ref{fig:2}a is shown. The spacing between the peaks is given by $\Delta V_\mathrm{CH} = 0.2~$V. For the addition of a single charge carrier, this corresponds to a capacitance between the dot and the channel gate of $C_\mathrm{CH} = e/\Delta V = 0.8~$aF, in rough agreement with the capacitance of $C_\mathrm{CH} = 1.3~$aF extracted using the gate voltage spacing of the Landau levels in Fig.~\ref{fig:1}b. For the latter estimate a dot area of $A = 0.02~\mu$m$^2$ (the lithographic size of the dot) was assumed. The significant background signal indicates that, apart from the channel exhibiting Coulomb blockade, there is another conductive channel through the dot. The conductive background decreases with increasing magnetic field (see Fig.~\ref{fig:2}a), but above $B = 3~$T the tunnel coupling to the reservoirs also gets weaker, because the distance between edge channels increases. Therefore we do not obtain a clearer signal at higher magnetic fields. The line cut shows a series of alternating low and high current peaks. This is reminiscent of the work on GaAs dots by Baer et al. \cite{Baer2013}, in which the pattern was explained by a different tunnel coupling to the reservoirs for the inner and outer edge channel present in the dot.

The conductance at $B = 2~$T as a function of density in the bulk $n_\mathrm{BG}$ and density in the channel $n_\mathrm{CH}$ (extracted from Fig.~\ref{fig:1}b) is shown in Fig.~\ref{fig:2}d. In Fig.~\ref{fig:2}e the same measurement is shown with a smoothened background subtracted. From the data it is apparent that the resonances (marked by dashed lines) are independent of the density in the bulk of the device and only depend on the density inside the dot for the entire range of the measurement. This is in agreement with our expectation, since the density in the bulk is much higher than the density in the dot. It also shows that the dot is clearly located inside the channel, making a disorder induced dot as in Ref.~\cite{Tovari2016b} highly unlikely.

\begin{figure}
\centering
\includegraphics[width=0.5\textwidth]{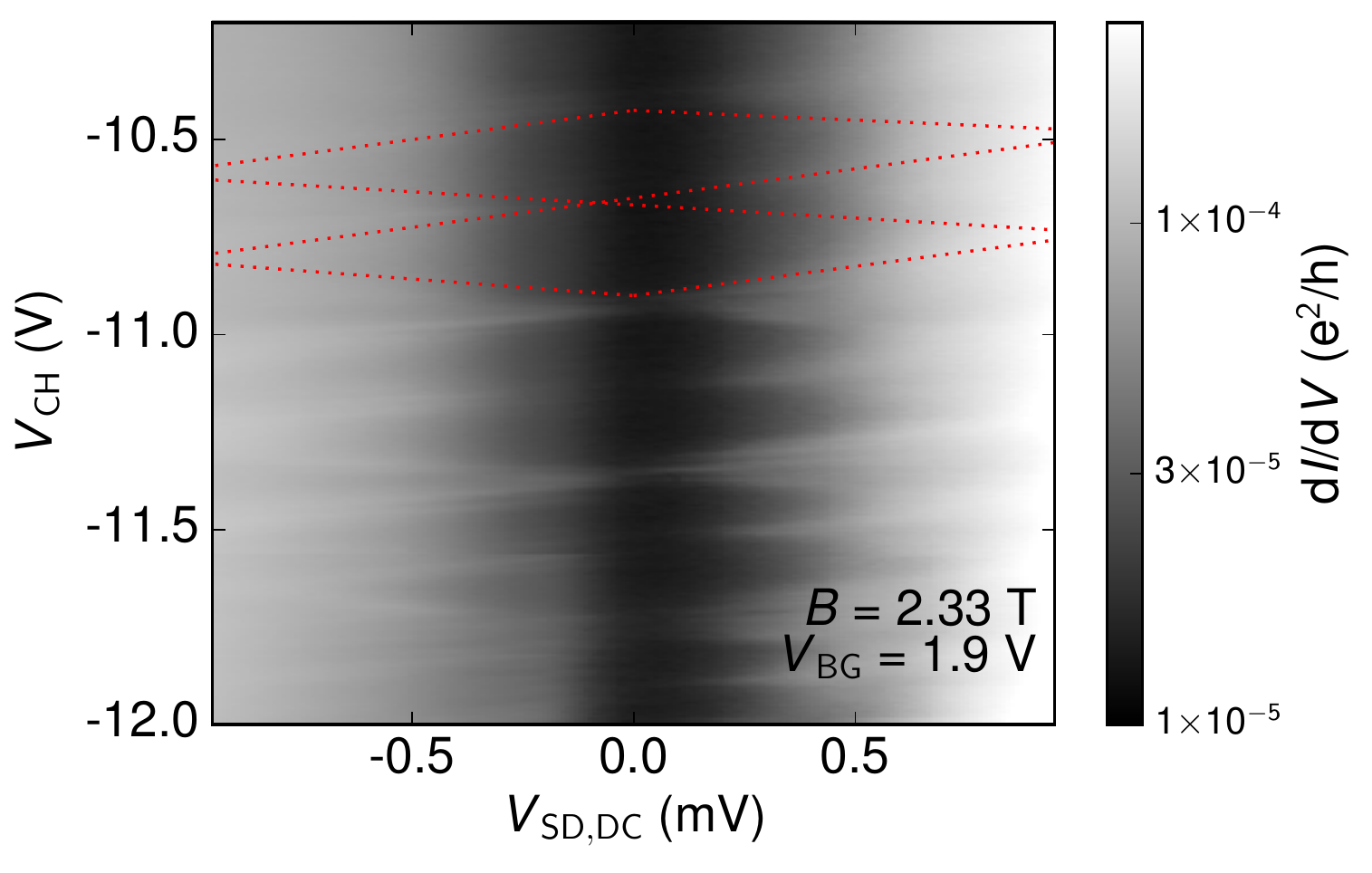}
\caption{Bias spectroscopy. The differential conductance as a function of source drain bias and channel gate voltage measured at $B~=~2.33~$T and V$_\mathrm{BG}~=~1.9~$V (dotted line in Fig.~\ref{fig:2}b). Pairs of Coulomb diamonds are visible, as indicated by the dotted red lines. They show an energy scale of roughly 1 meV. }
\label{fig:3}
\end{figure}

Finite bias measurements were performed to determine the energy spacing of the levels in the quantum dot. The differential conductance as a function of dc bias and channel gate voltage (see Fig.~\ref{fig:3}) at $B = 2.33~$T shows pairs of Coulomb diamonds (see dotted red lines), with a charging energy on the order of 1 meV. By approximating the dot as a disk in an infinite medium with a self capacitance of $C = 8 \epsilon \epsilon_0 r$ , we can extract a radius of $r = 500~$nm. This is an upper bound for the dot radius, since the presence of closeby metal gates alters the capacitance of the dot significantly \cite{Kouwenhoven1997}.

\section{Conclusion}
We realized an edge channel quantum dot in bilayer graphene making use of $p$-$n$ interfaces. Around B = 2 T, the device shows Coulomb blockade. A scaled up version of the device discussed in this work could be used as an Aharanov-Bohm interferometer \cite{Zhang2009a}. In the light of the recent observations of robust even denominator fractional quantum Hall states in bilayer graphene \cite{Zibrov2017,Li2017a}, it could be an interesting geometry to study the statistics of the quasiparticles of these states \cite{Stern2006}.

\section*{Acknowledgements}
We thank Tobias Kr{\"a}henmann, Szymon Hennel and Marc R{\"o}{\"o}sli for fruitful discussions. We thank Thomas Maurer for his assistance during the experiments. We acknowledge financial support from the European Graphene Flagship and the Swiss National Science Foundation via NCCR Quantum Science and Technology. Growth of hexagonal boron nitride crystals was supported by the
Elemental Strategy Initiative conducted by the MEXT, Japan and JSPS
KAKENHI Grant Numbers JP15K21722.

\section*{References}

\bibliography{dot}
\end{document}